\documentclass[a4paper, twoside, reqno, 12pt, dvips]{amsart}
\RequirePackage{fix-cm} 

\usepackage{fixltx2e}     

\usepackage[english]{babel}
\usepackage[latin1]{inputenc}

\usepackage{eucal}
\usepackage{esint}
\usepackage{amsgen}
\usepackage{amsthm}
\usepackage{xspace}
\usepackage{amssymb}
\usepackage{amsmath}
\usepackage{amsfonts}
\usepackage{verbatim}
\usepackage{mathtools}
\usepackage{mathrsfs} 

\usepackage[nice]{nicefrac}

\usepackage{a4wide}

\usepackage{microtype}
\usepackage[bf, up, hang]{caption}

\usepackage{indentfirst}
\usepackage{graphicx}
\usepackage{subfigure}
\usepackage[section]{placeins}
\usepackage{psfrag}

\usepackage{epsfig}

\usepackage{color}
\definecolor{oneblue}{rgb}{0.0, 0.0, 0.85}
\definecolor{darkgrey}{rgb}{0.273, 0.281, 0.30}

\usepackage{xcolor}
\definecolor{lightgray}{gray}{0.9}

\usepackage[colorlinks,
            urlcolor=oneblue,
            linkcolor=oneblue,
            citecolor=oneblue,
            bookmarksopen=false,
            pagebackref]{hyperref}

\usepackage[compact]{titlesec}

\titleformat{\section}{\normalfont\Large\bfseries\sffamily\center\color
{darkgrey}}{\thesection.}{0.5em}{}{}
\titleformat{\subsection}{\normalfont\large\bfseries\sffamily\color{darkgrey}}
{\thesubsection.}{0.4em}{}{}
\titleformat{\subsubsection}{\normalfont\normalsize\bfseries\sffamily\color
{darkgrey}}{\thesubsubsection.}{0.3em}{}{}

\titlespacing*{\section}{1.0em}{1.0em}{0.8em}[0em]
\titlespacing*{\subsection}{1.0em}{1.0em}{0.8em}[0em]
\titlespacing*{\subsubsection}{1.0em}{0.7em}{0.6em}[0em]

\usepackage{titletoc}

\contentsmargin{0.0em}
\dottedcontents{section}[2.5em]{\addvspace{0.45em}\bfseries}{1.0em}{0pc}
\dottedcontents{subsection}[4.5em]{}{2.6em}{0pc}
\dottedcontents{subsubsection}[5.5em]{}{3.0em}{0pc}

\usepackage{fancyhdr}
\usepackage{lastpage}

\newcommand*\Title{Visco-potential flows in Electro-hydrodynamics}
\newcommand*\Authors{M.~Hunt \& D.~Dutykh}

\numberwithin{equation}{section}

\newtheorem{remark}{Remark}

\newcommand{\E}{\mathbf{E}}
\newcommand{\T}{\mathbf{T}}

\newcommand{\n}{\mathbf{n}}
\newcommand{\R}{\mathbb{R}}
\newcommand{\ud}{\mathrm{d}}
\newcommand{\ue}{\mathrm{e}}
\newcommand{\ui}{\mathrm{i}}
\renewcommand{\phi}{\varphi}
\newcommand{\vi}{\mathbf{i}}
\newcommand{\vj}{\mathbf{j}}
\newcommand{\vk}{\mathbf{k}}
\newcommand{\A}{\mathcal{A}}

\renewcommand{\u}{\mathbf{u}}
\newcommand{\Zero}{\mathbf{0}}
\newcommand{\eps}{\varepsilon}
\renewcommand{\O}{\mathcal{O}}

\DeclareMathOperator{\sgn}{sgn}

\newcommand{\scal}{\boldsymbol{\cdot}}
\newcommand{\grad}{\boldsymbol{\nabla}}
\newcommand{\pd}[2]{\frac{\partial #1}{\partial\/ #2}}
\renewcommand{\div}{\grad\scal}
\renewcommand\Re{\operatorname{Re}}

\begin{document}

\title[\Title]{Visco-potential flows in electrohydrodynamics}

\author[M.~Hunt]{Matthew Hunt$^*$}
\address{Department of Mathematics, Faculty of Mathematical and Physical 
Sciences, University College London, Gower Street, London, WC1E  6BT, UK}
\email{mat@hyperkahler.co.uk}
\urladdr{http://hyperkahler.co.uk/}
\thanks{$^*$ Corresponding author}

\author[D.~Dutykh]{Denys Dutykh}
\address{University College Dublin, School of Mathematical Sciences, Belfield,
 Dublin 4, Ireland \and LAMA, UMR 5127 CNRS, Universit\'e de Savoie, Campus
 Scientifique, 73376 Le Bourget-du-Lac Cedex, France}
\email{Denys.Dutykh@ucd.ie}
\urladdr{http://www.denys-dutykh.com/}

\begin{abstract}

In this study we consider the problem of the interface motion under the 
capillary--gravity and an external electric force. The infinitely deep fluid 
layer is assumed to be viscous, perfectly conducting and the flow to be 
incompressible. The weak viscous effects are introduced using the 
Helmholtz--Leray decomposition and the visco-potential flow approach. The 
electric charge distributions above and on the free surface are considered.
Finally, we derive some linearized analytical solutions for the free 
surface elevation shape under the localised pressure distribution and the 
combined action of the forces mentioned hereinabove.

\bigskip
\noindent \textbf{\keywordsname:} electro-hydrodynamics; free surface flows; 
viscous dissipation; potential flow
\end{abstract}

\maketitle

\tableofcontents

\section{Introduction}
Liquid thin films are a common occurance in the fields of biology and 
engineering under the guise of coating flows, and have been subject to 
intensive study. Instabilities and the ensuing dynamics associated to the liquid 
film can by caused by a number of effects including interfacial 
instabilities due to surface tension variations (e.g. Marangoni instabilities), 
gravitational instabilities (such as Rayleigh--Taylor) as well as instabilities 
caused by external fields like electric or magnetic ones. Instabilities of the 
interface can also be due to topography or a moving pressure distribution (as it 
is considered in this study), but an external electric field may lead to the 
stabilization of the interface. More precisely, depending on the asymptotic form 
of the electric field in question, its effect can be either stabilizing 
\cite{Tseluiko2009} or destablizing \cite{Taylor1965}.

The study of electro-capillary waves was first initiated by G.I.~Taylor 
\emph{et al.}\cite{Taylor1965} for 3D waves with a prescribed disturbance of 
the interface. On the other hand, in modern investigations the disturbance is 
determined according to an applied external electric field with the free 
surface profile is calculated by solving the electro-hydrodynamic formulation 
\cite{Papageorgiou2004}. The work carried out by Hunt \cite{Hunt2013} focused on 
forced waves in electrohydrodynamics using the methods set out in 
\cite{Papageorgiou2004, Vanden-Broeck2010} to examine linear and weakly 
nonlinear free surface flows in 2D and weakly 3D (Kadomtsev-Petviashvili-type 
models). The method proposed in the present study takes into account the viscous effects \cite{Tseluiko2006, Tseluiko2007, Tseluiko2008, 
Tseluiko2009} by keeping the \emph{simplicity} of the potential flow 
approach. For the sake of the clarity of the exposition, we illustrate this 
method in 2D inifinite depth case.

The theory of visco-potential flows probably originates from the pioneering 
work of J.~Boussinesq (1895) \cite{Boussinesq1895} who estimated the water 
wave amplitude decay due to the effect of viscosity in the linear 
approximation. Then, this research has been continued in the 70's in the context 
of nonlinear long wave models \cite{Ott1970, KM}. Later, Kit \& Shemer (1989) 
\cite{Kit1989} developed a theoretical model which allows the estimation of the 
wave energy dissipation which included the friction effect at the bottom and 
lateral walls in a rectangular wave tank. The potential flows of viscous 
fluids were also thoroughly investigated theoretically by D.~Joseph and his 
collaborators \cite{Joseph1994, Joseph2006}. The visco-potential formulation 
in the deep water case was derived by Dias \emph{et al.} (2008) \cite{Dias2007}
and generalized to the finite depth case in \cite{Liu2004} and later 
in \cite{DutykhDias2007, Dutykh2008a}. This formulation was validated 
experimentally and numerically for the practically important case of solitary 
wave propagation in \cite{Liu2006}. More recently, the damping rates for 
various dissipative operators were investigated numerically in \cite{Chehab2011,
Sadaka2013}. Finally, the asymptotic long time behaviour of some 
visco-potential models were found and justified analytically in \cite{CG, 
Chen2010, Goubet2010}.

The present manuscript is organized as follows. After a brief introduction, in 
Section~\ref{sec:model} we present the derivation of the electro-hydrodynamic 
visco-potential formulation. An analytical solution to the linearized 
formulation is shown in Section~\ref{sec:res} for several values of problem 
parameters. A particular case of the electro-hydrodynamic problem where the 
charge is distributed on the interface is considered in 
Appendix~\ref{sec:charge}. Finally, the main conclusions and perspectives of 
this study are outlined in Section~\ref{sec:concl}.

\section{Mathematical Model}\label{sec:model}

\begin{figure}
\centering
\includegraphics[scale=0.75]{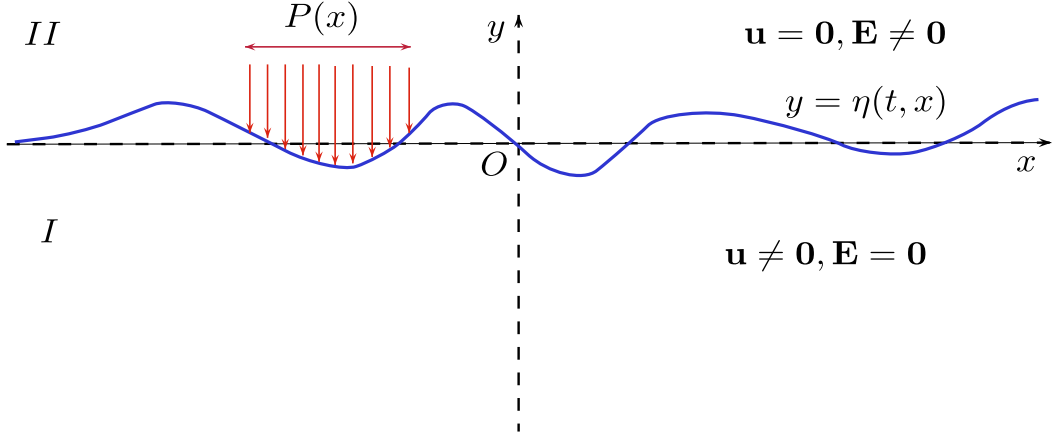}
\caption{\small\em Sketch of the physical domain considered in this study.}
\label{fig:sketch}
\end{figure}

Consider an infinitely deep channel of a perfectly conducting weakly viscous 
and incompressible fluid (referred to as the heavy fluid) and is in the region 
$\left\{(x,y)|\, x\in\R,\, y\leqslant \eta(x,t)\right\}$. The flow is assumed 
to be exactly two-dimensional. A Cartesian coordinate system $(x, y)$ is 
introduced, with $y$ pointing vertically upwards. The interface between two 
fluids is given by $y = \eta(x,t)$. The sketch of the physical domain is given 
on Figure~\ref{fig:sketch}. Below we will adopt the so-called free surface 
assumption and the light fluid will enter into equations only through the 
electric forces.

The velocity field in the heavy fluid in region $I$ is then given as $\u = u\vi 
+ v\vj + 0\vk$. We assume that there is an electric field in region $II$ 
$\left\{(x,y)|\, x\in\R,\, y > \eta(x,t)\right\}$. The electric field $\E$ 
satisfies the equations $\grad\times\E = \Zero$ and, thus, can be written in 
the following potential form:
\begin{equation*}
  \E \equiv -\grad V.
\end{equation*}
Since there are no free charges in the light fluid, the Gauss's law states 
that $\div\E = 0$, and hence the governing equation for the electric 
potential reads:
\begin{equation*}
  \grad^2 V \equiv \pd{^2 V}{x^2} + \pd{^2 V}{y^2} = 0 \qquad y > \eta(x,t).
\end{equation*}
This equation is completed by appropriate boundary conditions. At infinity we 
require the following asymptotic behaviour of the solution $V(x,y)$:
\begin{equation*}
  V(x,y) \to -E_{0}y \qquad \textrm{as}\qquad y\to +\infty
\end{equation*}
The usual linear incompressible Navier-Stokes equations are used and use the Helmholtz decomposition \cite{Dias2007,
 DutykhDias2007} on the velocity vector $\mathbf{u}=(u,v,0)$
\begin{equation}\label{eq:decomp}
  \u = \grad\phi + \grad\times\A,
\end{equation}
with $\A =(0,0,\psi)$. 
The idea of the approach taken here is to say that the motion is mainly potential flow but with a small non-potential part, represented by the function $\psi$ i.e. $\|\psi\|_{L^{2}} 
\ll \|\varphi\|_{L^{2}}$.\footnote{The $L^{2}$ norn is given by: $$\|f\|_{L^{2}}=\left(\int_{\mathbb{R}^{2}}|f(x,y)|^{2}dxdy\right)^{\frac{1}{2}}$$} Along with a small viscosity, $\nu$.
The decomposition of the velocity is inserted into the linear Navier-Stokes equations to end up with two separate equations for $\varphi$ and $\psi$
\begin{equation*}
  \pd{\psi}{t} = \nu\grad^2\psi ,\quad\nabla^{2}\varphi=0
\end{equation*}
Moreover, in 2D the velocity field $\u$ can be simply expressed as:
\begin{equation*}
  u = \pd{\phi}{x} + \pd{\psi}{y}, \quad v = \pd{\phi}{y} - \pd{\psi}{x}.
\end{equation*}
The use of decomposition for the velocity and the governing equation for $\psi$ yields a Bernoulli equation in the usual way: 
\begin{equation}\label{eq:vb}
  \pd{\phi}{t} + \frac{p_{I}}{\rho} + gy = C.
\end{equation}
\subsection{Boundary Conditions}

In constrast to some previous studies which considered the linear case, in 
this section the derivations of the fully nonlinear equations will be presented.

The capillary and electric effects come into the problem through the Young--
Laplace equation at the interface between two media:
\begin{equation}\label{eq:yl}
  \bigl[\hat{\n}\cdot\T\cdot\hat{\n}\bigr]_1^2 = \sigma\div\hat{\n},
\end{equation}
where the brackets $[\cdot]$ denotes evaluation at each side of the interface $y = \eta(x,
t)$ and the unit normal $\hat{\n}$ (pointing from the lower fluid to the upper 
fluid) and unit tangent vector are given by
\begin{equation}\label{eq:normal}
  \hat{\n} = \frac{(-\eta_x, 1)}{\sqrt{1 + \eta_x^2}},\quad\hat{\mathbf{t}}=\frac{(1,\eta_{x})}{\sqrt{1+\eta_{x}^{2}}}
\end{equation}
The stress tensor take on two different forms depending upon which region is 
under consideration. In region I, 
\begin{equation*}
  T_{ij} = -p\delta_{ij} + \tau_{ij},
\end{equation*}
where $p$ is the pressure as defined above and $\delta_{ij}$ is the Kronecker
delta symbol. and the tensor $\tau_{ij}$ corresponds to the viscosity.
The stress tensor in region $II$ is given by
\begin{equation*}
  T_{ij} = -P\delta_{ij} + \Sigma_{ij},
\end{equation*}
where $P$ is the pressure distribution \emph{on} the interface and $\Sigma_{ij}$ 
corresponds to the electric field in region $II$. Two constitutive components 
of the viscous and electric stresses are correspondingly:
\begin{itemize}
	\item $\tau_{ij} = \mu\bigl(\partial_j u_i + \partial_i u_j
	\bigr)$,
	\item $\Sigma_{ij} = \eps_p\Bigl(E_i E_j - \frac{1}{2}
	\delta_{ij} E_k E_k\Bigr)$,
\end{itemize}
where $\mu = \rho\nu$ is the dynamic viscosity and $\eps_p$ is the electric permittivity.                                                                                 
Recall that the Young--Laplace condition \eqref{eq:yl} involves the 
projection of the stresses onto the normal direction $\hat{\n}$ to the 
interface. In order to avoid cumbersome expressions, we will perform the 
computations of $\hat{\n}\cdot\T\cdot\hat{\n}$ by parts. The viscous part of 
the stress tensor $\T$ gives:
\begin{multline*}
  \hat{\n}\scal\boldsymbol{\tau}\scal\hat{\n} = \hat{n}_i\tau_{ij}\hat{n}_j = 
  \hat{n}_1^2\tau_{11} + \hat{n}_2^2\tau_{22} + 2\hat{n}_1\hat{n}_2\tau_{12} = 
  \\ \frac{\nu}{1+\eta_x^2}\Bigl[\eta_x^2(\phi_{xx} + \psi_{xy}) - 2\eta_x (
  2\phi_{xy} + \psi_{yy} - \psi_{xx}) + 2(\phi_{yy} - \psi_{xy})\Bigr].
\end{multline*}
The same expansion can be also done for the Faraday stress tensor $\Sigma$:
\begin{equation*}
  \hat{\n}\scal\Sigma\scal\hat{\n} = \hat{n}_1^2\Sigma_{11} 
  + \hat{n}_2^2\Sigma_{22} - 2\hat{n}_1\hat{n}_2\Sigma_{12},
\end{equation*}
where
\begin{equation*}
  \Sigma_{11} = \frac{\eps_p}{2}(V_x^2 - V_y^2), \quad
  \Sigma_{12} = \eps_p V_x V_y, \quad
  \Sigma_{22} = -\frac{\eps_p}{2}(V_x^2 - V_y^2).
\end{equation*}
Where the approximation $\|\psi\|_{L^{2}}\ll\|\varphi\|_{L^{2}}$ was used along with a small viscosity.
\begin{multline*}
  \pd{\phi}{t} +\frac{1}{2}|\nabla\varphi |^{2}+ g\eta +  \frac{P}{\rho} - \frac{1}{\rho(1 + \eta_x^2)}\Bigl[
  \eta_x^2\Sigma_{11} + 2\eta_x\Sigma_{12} + \Sigma_{22}\Bigr]
  + \frac{\nu}{1 + \eta_x^2}\Bigl(\eta_x^2\phi_{xx} \\
  -2\eta_x\phi_{xy} + 2\phi_{yy}\Bigr) 
  = \frac{\sigma}{\rho}\frac{\eta_{xx}}{(1 + \eta_x^2)^{3/2}}, \quad \mbox{on} 
  \quad y = \eta(x,t).
\end{multline*}


\noindent
The free surface equation is modified in the presence of a small viscosity which is on of  
the results of the visco-potential flow theory \cite{Liu2004, Dias2007, 
DutykhDias2007}: 
\begin{equation*}
  \pd{\eta}{t} = \pd{\phi}{y} + 2\nu\pd{^2\eta}{x^2}.
\end{equation*}

\subsection{Fully Nonlinear Formulation}\label{sec:full}

Now we can write down the set of fully nonlinear equations which govern the 
motion of the viscous fluid under the action of an exterior electric force:
\begin{equation*}
  \pd{^2\phi}{x^2} + \pd{^2\phi}{y^2} = 0,
  \qquad -\infty < y \leqslant \eta(x,t)
\end{equation*}
\begin{equation*}
  \pd{^2V}{x^2} + \pd{^2V}{y^2} = 0,
  \qquad \eta(x,t) < y < +\infty 
\end{equation*}
\begin{equation*}
  \pd{\eta}{t} + \pd{\phi}{x}\pd{\eta}{x} =\pd{\phi}{y} + 2\nu\pd{^2\eta}{x^2},
  \qquad y = \eta(x,t)
\end{equation*}
\begin{multline*}
  \pd{\phi}{t} + \frac{1}{2}|\nabla\varphi |^{2}+g\eta +  \frac{P}{\rho} - \frac{1}{\rho(1 + \eta_x^2)}\Bigl[
  \eta_x^2\Sigma_{11} + 2\eta_x\Sigma_{12} + \Sigma_{22}\Bigr]
  \\ + \frac{\nu}{1 + \eta_x^2}\Bigl(\eta_x^2\phi_{xx} 
  -2\eta_x\phi_{xy} + 2\phi_{yy}\Bigr) = \frac{\sigma}{\rho}
  \frac{\eta_{xx}}{(1 + \eta_x^2)^{3/2}}, \quad \mbox{on} \quad y = \eta(x,t).
\end{multline*}
\begin{equation}\label{eq:scharge}
  \pd{V}{x} + \pd{\eta}{x}\pd{V}{y} = 0, \quad y = \eta(x,t)
\end{equation}
\begin{equation*}
  V \rightarrow -E_0 y \qquad y\rightarrow +\infty,
\end{equation*}
\begin{equation*}
  \pd{\phi}{y} \rightarrow 0, \qquad y\rightarrow -\infty
\end{equation*}
This set of equations will be used below to derive some linearised solutions 
to the electro-hydrodynamic problem of weakly viscous fluids. A particular 
case of this problem when the charge is distributed on the interface is 
discussed in Appendix~\ref{sec:charge}.

\begin{remark}
The formulation presented above was done in deep water approximation. The 
generalization to the finite depth case can be done relatively easily. The 
usual bottom impermeability condition is modified to include a nonlocal term 
in time which is due to the presence of a boundary layer at the bottom. the 
details can be found in \cite{Liu2004, DutykhDias2007, Dutykh2008a}. For the 
same reasons, the dispersion relation of the visco-electro-hydrodynamic problem 
coincides with the classical visco-potential formulation \cite{Dutykh2008a,
Dutykh2008b}.
\end{remark}

\subsection{Linear Wave Theory on a Uniform Flow}
\noindent
The method of derivation in this case for infinite deph is given in \cite{Vanden-Broeck2010} section 4.2.2.
It is assumed that there is a uniform flow with speed $U$, and choose 
the asymptotic expansion as follows:
\begin{eqnarray*}
  \phi &=& Ux + \eps\phi_1 + o(\eps) \\
  \eta &=& \eps\eta_1 + o(\eps) \\
     V &=& -E_0 y + \eps V_1 + o(\eps) \\
     p &=& \eps p_1 + o(\eps)
\end{eqnarray*}
The linearisation of the system of equations presented in 
Section~\ref{sec:full} reads:
\begin{eqnarray}
  \pd{^2\phi_1}{x^2} + \pd{^2\phi_1}{y^2} &=& 0
  \quad -\infty <y \leqslant 0 \nonumber \\
  \pd{^2V_1}{x^2} + \pd{^2V_1}{y^2} &=& 0
  \quad 0 \leqslant y < +\infty \nonumber \\
  U\pd{\eta_1}{x} &=& \pd{\phi_1}{y} + 2\nu\pd{^2\eta_1}{x^2}
  \quad y = 0 \nonumber \\ 
  U\pd{\phi_1}{x} + \frac{p_1}{\rho} + g\eta_1 
  + \frac{\eps_p E_0^2}{\rho}\pd{V_1}{y} + 2\nu\pd{^2\phi_1}{y^{2}} &=& 
  \frac{\sigma}{\rho}\pd{^2\eta_1}{x^2} \quad y = 0 \label{eq:b} \\
  \pd{V_1}{x} &=& E_0\pd{\eta_1}{x} \nonumber \\ 
  \pd{\phi_1}{y} &\rightarrow& 0 \quad y\rightarrow -\infty \nonumber \\
  \pd{V_1}{y}    &\rightarrow& 0 \quad y\rightarrow +\infty \nonumber
\end{eqnarray}
The obtained linear system will be understood uising the Fourier analysis:
\begin{equation*}
\begin{multlined}
  \phi_1(x) = \frac{1}{2\pi}\int_{\R}\hat{\phi}_1\ue^{\ui kx}\,\ud k, \quad 
  V_1(x)    = \frac{1}{2\pi}\int_{\R}\hat{V}_1\ue^{\ui kx}\,\ud k, \\
  \eta_1(x) = \frac{1}{2\pi}\int_{\R}\hat{\eta}_1\ue^{\ui kx}\,\ud k, \quad 
  p_1(x)    = \frac{1}{2\pi}\int_{\R}\hat{p}_1\ue^{\ui kx}\,\ud k.
\end{multlined}
\end{equation*}
The solutions for $\phi_1(x,y)$ and $V_1(x,y)$ can be easily obtained:
\begin{equation*}
  \phi_1(x,y) = \frac{1}{2\pi}\int_{\R}A(k)\ue^{|k|y}\ue^{\ui kx}\,\ud k, \quad
     V_1(x,y) = \frac{1}{2\pi}\int_{\R}B(k)\ue^{-|k|y}\ue^{\ui kx}\,\ud k.
\end{equation*}
By satisfying the boundary conditions one can find the unknown functions $A(k)$
and $B(k)$:
\begin{equation*}
  A(k) = (\ui U + 2\nu k)\sgn(k)\hat{\eta}_1(k), \qquad
  B(k) = E_0\hat{\eta}_{1}(k).
\end{equation*}
Inserting all the elements into equation \eqref{eq:b} allows us to find the 
Fourier transform of the free surface elevation:
\begin{equation*}
  \hat{\eta}_1(k) = \frac{\hat{p}_1(k)}{\rho U^2}\Bigl[k\sgn(k) 
  -\frac{4\ui k^2\nu\sgn(k)}{U} -\frac{g}{U^2}+\frac{\eps_p E_0^2}{\rho U^2}|k| 
  -\frac{4\nu^2 k^3\sgn(k)}{U^2} - \frac{\sigma}{\rho U^2}k^2\Bigr]^{-1}.
\end{equation*}
For brevity denote $s_{k}=\sgn(k)$ and note that $ks_{k}=|k|$. Following section 4.2.2 in \cite{Vanden-Broeck2010}, a Gaussian distribution distribution is used:
\begin{equation*}
  p_1(x) = \frac{\rho U^2}{2}\ue^{\frac{-5g^2 x^2}{U^4}}.
\end{equation*}
The expression for the free surface elevation in the physical space under this 
pressure distribution is then given by:
\begin{equation}\label{eq:eta11}
  \eta_1(x) = \frac{U^2}{4g\sqrt{5\pi}}\Re\int_{\R}
  \frac{\ue^{-\frac{k^2 U^4}{20g^2}}e^{\ui kx}}{|k| - \frac{4\ui\nu k|k|}{U}
  -\frac{g}{U^2} + \frac{\eps_p E_0^2 |k|}{\rho U^2} 
  - \frac{4\nu^2 k^2|k|}{U^2} - \frac{\sigma k^2}{\rho U^2}}\,\ud k.
\end{equation}
This expression can be further simplified by choosing new dimensionless 
variables:
\begin{equation*}
  l := \frac{U^2}{g}k, \quad 
  \hat{x} := \frac{g}{U^2}x, \quad
  \zeta := \frac{g}{U^2}\eta_1.
\end{equation*}
With these new variables the solution \eqref{eq:eta11} becomes:
\begin{equation}\label{eq:sol}
  \zeta(\hat{x}) = \frac{1}{4\sqrt{5\pi}}\Re\int_{\R}
  \frac{\ue^{-\frac{l^2}{20}}e^{\ui l\hat{x}}}{(1 + \beta)|l| - 1 
  - \ui\delta l|l| - \gamma l^2|l| - \alpha l^2}\, \ud l,
\end{equation}
where
\begin{equation*}
  \alpha := \frac{\sigma g}{\rho U^4}, \quad
  \beta := \frac{\eps_p E_0^2}{\rho U^2}, \quad
  \gamma := \frac{4\nu^2 g^2}{U^6}, \quad
  \delta := \frac{4\nu g}{U^3}.
\end{equation*}
For instance, we can notice that the coefficient $\gamma = \O(\delta^2) \ll 1$ and, thus, can be neglected. The integrand in 
\eqref{eq:sol} is complex. The complex part plays the role of the Rayleigh 
dispersion as mentioned in \cite{Vanden-Broeck2010}. From an analytical point of 
view, the complex terms remove the singularity. Consequently, formula 
\eqref{eq:sol} will always define a valid solution in $L^{2}$ due to the 
complex denomenator. It is expected also that $\zeta (\hat{x})\rightarrow 0$ 
as $\hat{x} \rightarrow \pm\infty$.

\section{Results}\label{sec:res}
\begin{figure}
\centering
\includegraphics[scale=0.75]{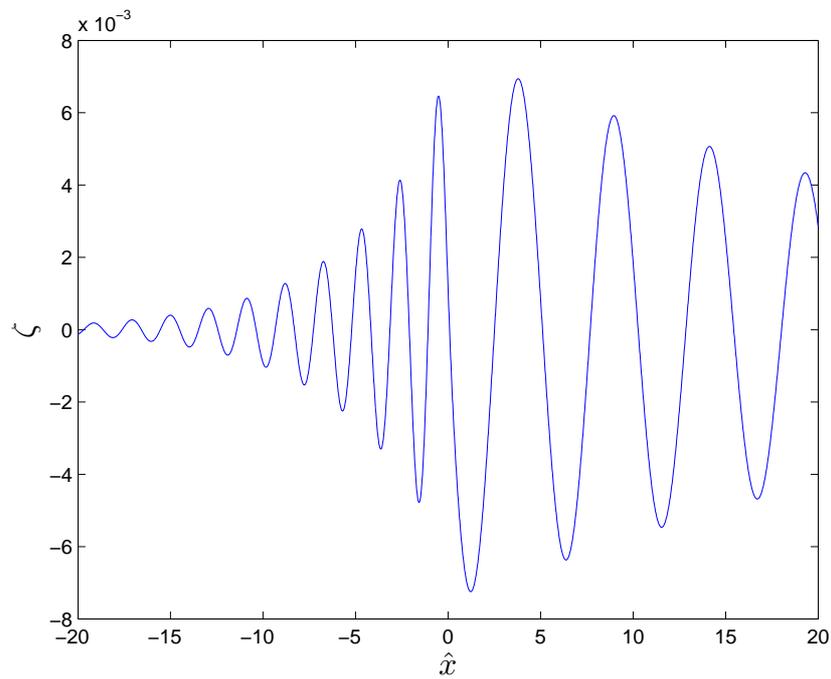} 
\caption{\small\em Typical free surface profile predicted by solution 
\eqref{eq:sol}.}
\label{fig:vfs}
\end{figure}

\noindent
Comparisons of \eqref{eq:sol} can be made with previous established results in the literature, taking $\beta =\gamma =\delta =0$ reduces to the case in \cite{Vanden-Broeck2010} and likewise setting $\gamma =\delta =0$ reduces to the case in \cite{Hunt2013}. 
In order to illustrate the analytical result derived in the previous Section, 
we plot on Figure~\ref{fig:vfs} the free surface elevation shape predicted 
analytically by solution \eqref{eq:sol}. The parameters were chosen to be 
$\alpha = 0.3$, $\beta = 0.15$ and $\delta = 0.01$. The profile for these base 
values is given in Figure~\ref{fig:vfs}.

\begin{figure}
  \centering
  \includegraphics[width=0.99\textwidth]{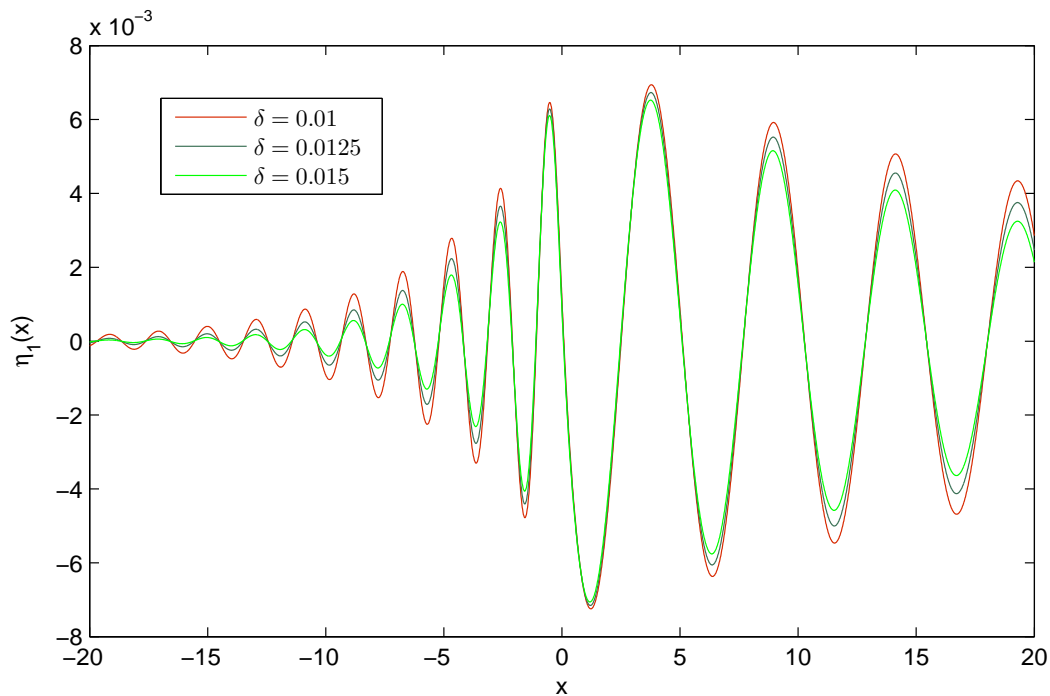}
  \caption{\small\em Free surface elevation shapes for various values of the 
  parameter $\delta$.}
  \label{fig:paramd}
\end{figure}

When $\delta =0$, there is a possibility that there are two zeros in the denominator of \eqref{eq:sol} if the values are taken above the minimum. The weakly viscous terms act as a kind of variable Rayleigh dispersion term which is why there is more damping than is normally seen with the inviscid case.
In order to compare the viscous case with previous cases, the case in \cite{Hunt2013} is given by:

\begin{figure}
\centering
\includegraphics[scale=0.75]{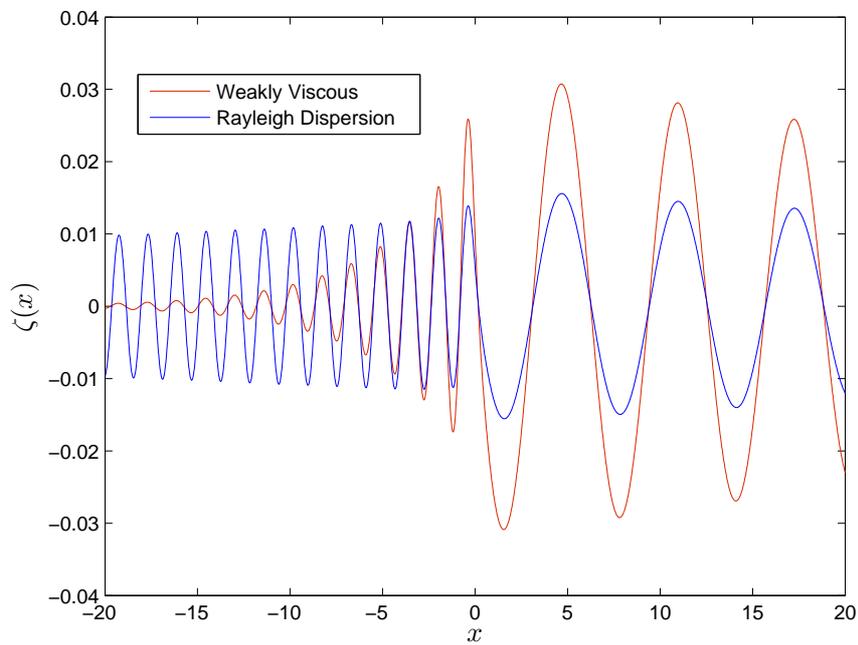}
\caption{Comparison of Non-Viscous Waves with Rayleigh Dispersion and Viscous Waves}
\label{comparison}
\end{figure}
As can be seen quite clearly, even a small viscosity has a large impact on wave of higher frequency by damping it severly and increases the amplitude of the lower frequency whilst also damping the waves. It is also interesting that the addition of viscosity doesn't affact the wavelength of the short and long waves. In order to illustrate the free surface elevation depends on the parameter $\beta$ (which measures the relative importance of intertia to the electric force), we represent in figure X the same solution (\eqref{eq:sol}) for several values of $\beta$
\begin{figure}
\centering
\includegraphics[scale=0.75]{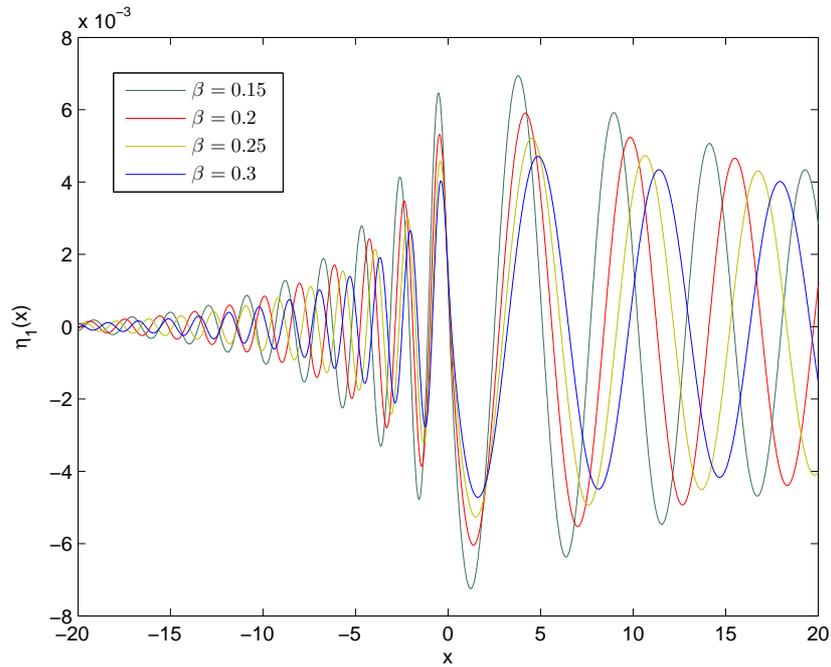}
\caption{Free Surface Profiles for Various Values of $\beta$}
\label{comparison}
\end{figure}
The same result is plotted on Figure~\ref{fig:paramd} also for the parameter 
$\delta$, which measures the relative magnitude of dissipative effects. The 
increase in parameters leads the slight decrease of the amplitude, as it can be 
expected from the analytical solution.

\section{Conclusions \& Perspectives}\label{sec:concl}

In this study we considered the problem of a free surface flow description in 
the presence of electric forces. Moreover, the fluid is assumed to be weakly 
viscous. By using the visco-potential flow theory, the classical potential 
formulation was modified to take into account weak dissipative effects. The 
derivations presented in this study were performed in the deep water 
approximation. However, the generalization to the finite depth case does not 
represent any major difficulties. 

The free surface shape in the presence of a uniform current and a localized 
interfacial pressure distribution was computed analytically. The dependence of 
this shape on two important dimensionless parameters was studied numerically 
as well.

This study opens a certain number of directions for future investigations. For 
instance, the present analysis was only linear. Nonlinearities have to be 
taken into account as well as the finite depth effects. Moreover, long wave 
asymptotics can be also performed to investigate the shallow water regime.

\subsection*{Acknowledgments}
\addcontentsline{toc}{section}{Acknowledgments}

D.~\textsc{Dutykh} would like to acknowledge the support from ERC under the 
research project ERC-2011-AdG 290562-MULTIWAVE.

\appendix
\section{Electric charge at the interface}\label{sec:charge}

This section details the analysis of the problem where there is an interfacial
surface charge. In the inviscid case, as there were co shearing forces, there 
could be no dynamic charge on the surface and it was only possible to compute in 
\emph{induced} charge on the interface given by:
\begin{displaymath}
\Sigma_{Q}=\epsilon_{p}\frac{\partial V}{\partial n} 
\end{displaymath}
Then the definition leads to the same result as in the inviscid case, that the 
induced surface charge is just the Hilbert transform of the derivative of the 
free surface profile. For the \emph{dynamics} surface charge equation 
\eqref{eq:scharge} is replaced with two following equations \cite{Hammerton2013}:
\begin{equation*}
  \hat{\n}\cdot\bigl[\eps_p\E\bigr]_1^2 = q,
\end{equation*}
\begin{equation*}
  \pd{q}{t}+\grad_T\scal(q\mathbf{u}) + \hat{\n}\bigl[\sigma_q\E\bigr]_1^2 = 0.
\end{equation*}
The operator $\grad_T$ is the covariant derivative defined as
\begin{equation*}
  \grad_T = \grad - \hat{\n}\bigl(\hat{\n}\cdot\grad\bigr).
\end{equation*}
Below we will perform the linear analysis of this problem as well. The 
expansions in this case are given by:
\begin{eqnarray*}
 \phi &=& Ux + \eps\phi_1 + o(\eps) \\
 \eta &=& \eps\eta_1 + o(\eps) \\
    V &=& -E_0 x + \eps V_1 + o(\eps) \\
    q &=& \eps q_1 + o(\eps) \\
    p &=& \eps p_1
\end{eqnarray*}
The covariant derivative on the free surface with the normal defined by 
equation \eqref{eq:normal} is expressed as
\begin{equation*}
  \grad_T = \Bigl(\pd{}{x}, 0\Bigr) + \eps\Bigl(\pd{\eta_1}{x}\pd{}{y}, 
  \pd{\eta_1}{x}\pd{}{x}\Bigr) + o(\eps) \\
\end{equation*}
The equations reduce to the following \cite{Hammerton2013}:
\begin{equation*}
  U\pd{q_1}{x} + \sigma_q E_0\pd{\eta_1}{x} + \sigma_q \pd{V_1}{y} = 0
\end{equation*}
\begin{equation*}
  -\eps_p E_0\pd{\eta_1}{x} + \eps_p\pd{V_1}{y} = q_1
\end{equation*}
Two last equations can be combined into one:
\begin{equation*}
  -\eps_p E_0 U\pd{^2\eta_1}{x^2} + U\eps_p\pd{^2V_1}{x\partial y} + 
  \sigma_q E_0\pd{\eta_1}{x} + \sigma_q\pd{V_1}{y} = 0.
\end{equation*}
The rest of linearised equations in this case is exactly the same as in the 
previous section. The analytical expression for the linearized free surface 
elevation in the Fourier space can be derived in a similar way:
\begin{equation*}
  \hat{\eta}_1(k) = \Bigl[k\sgn(k) - \frac{4\ui\nu k^2\sgn(k)}U - \frac{g}{U^2} 
  + \frac{\eps_p E_0^2}{\rho U^2}k f(k) - \frac{\sigma}{\rho U^2}k^2\Bigr]^{-1}
  \frac{\hat{p}_1(k)}{\rho U^2}
\end{equation*}
Then, the free surface elevation in the physical space can be easily obtained:
\begin{equation*}
  \eta_1(x) = \frac{1}{2\pi \rho U^2}\int_{\R}
  \frac{\hat{p}_1(k)\ue^{\ui kx}}{k\sgn(k) - \frac{4\ui\nu k^2\sgn(k)}{U} 
  - \frac{g}{U^2} + \frac{\eps_p E_0^2}{\rho U^2} kf(k) 
  - \frac{\sigma}{\rho U^2}k^2}\,\ud k, \quad
  f(k) = \displaystyle{\frac{\eps_p Uk + \ui\sigma_q}
  {\sigma_q + \ui kU\eps_p}}
\end{equation*}
Using the same scalings and pressure form as before, the integral 
can be reduced to the following dimensionless form:
\begin{equation*}
  \zeta(\hat{x}) = \frac{1}{4\sqrt{5\pi}}\Re\int_{\mathbb{R}}
  \frac{\ue^{-\frac{l^2}{20}}\ue^{\ui\hat{x}l}}{|l| - \ui\delta l|l| - 1 
  + \beta l\tilde{f}(l) - \alpha l^2}\,\ud l,
\end{equation*}
where 
\begin{equation*}
  \tilde{f}=\frac{\mu l + \ui}{\ui\mu l + 1}, \qquad
  \mu = \frac{\eps_p g}{U\sigma_q},
\end{equation*}
and coefficients $\alpha$, $\beta$ and $\delta$ are defined as in the previous 
section.

\addcontentsline{toc}{section}{References}
\bibliographystyle{abbrv}
\bibliography{biblio}

\end{document}